\documentstyle[astrobib,epsfig]{mn}

\begin{document}

\def\bi#1{\hbox{\boldmath{$#1$}}}

\newcommand{\beq}{\begin{equation}}
\newcommand{\eeq}{\end{equation}}
\newcommand{\beqa}{\begin{eqnarray}}
\newcommand{\eeqa}{\end{eqnarray}}

\newcommand{\lexp}{\mathop{\langle}}
\newcommand{\rexp}{\mathop{\rangle}}
\newcommand{\rexpc}{\mathop{\rangle_c}}

\def\bi#1{\hbox{\boldmath{$#1$}}}

\title{Constraints on galaxy halo profiles  
from galaxy-galaxy lensing and Tully-Fisher/fundamental plane relations}

\author[U. Seljak]{
 U.~Seljak\thanks{E-mail: uros@feynman.princeton.edu} \\
 Department of Physics, Princeton University, Princeton, NJ 08544,
 USA}

\pubyear{2001}

\maketitle

\begin{abstract}
Observations of galaxy-galaxy lensing from 
Sloan Digital Sky Survey (SDSS) are combined with the Tully-Fisher 
and fundamental plane relations to derive constraints on galactic halo
profiles. 
We show that both for early and late type galaxies around 
$L_*$ the 
rotation velocity decreases 
significantly from its peak value at the optical radius 
to the virial radius $r_{200}$, $v_{\rm opt}/v_{200} \sim 1.8$ with about 20\% 
uncertainty.
Such a decrease is expected in models
in which the halo profile is very concentrated, so that it
declines steeper than isothermal at large radii. This large decrease  
can be explained as a result of both a concentrated dark matter profile
and a significant stellar contribution to the rotation velocity 
at the optical radii.
We model the stellar component with a thin rotationally supported disk or 
a Hernquist profile and use adiabatic dark matter response model to place 
limits on the halo concentration as a function of the stellar 
mass to light ratio. For reasonable values of the latter 
we find concentrations $c_{200} $ consistent with 
CDM predictions, suggesting  there is no evidence for low concentrations
for the majority of halos in the universe.
We also discuss the origin of Faber-Jackson relation $L \propto \sigma^4$
in light of $L\propto v_{200}^{2.5}$ relation found for early type galaxies
above $L_*$
from galaxy-galaxy lensing. This leads to a decrease in $v_{\rm opt}/v_{200}$ 
with luminosity above $L_*$, so that at $7L_*$ the ratio is 1.4.
This is expected from the fundamental plane relation as a result of 
a reduction in the baryonic contribution to the total mass at the optical 
radius and a decrease in optical to virial rotation velocity in dark matter
profile.
These results imply that relations such as Tully-Fisher and 
Faber-Jackson are not simply those between the
mass of dark matter halo and galaxy luminosity, but are also significantly 
influenced by the baryonic
effects on the rotation velocity at optical radii.

\end{abstract}

%\keywords{large-scale structure of universe}
\begin{keywords}
 cosmology: theory -- dark matter -- galaxies: haloes -- galaxies.
\end{keywords}

\section{Introduction}

The study of dark matter profiles around galaxies has been an active 
area since the original discovery of dark matter from the flat rotation 
curves in spirals (see \citeNP{2001ARA&A..39..137S} for a review).
These show that there must be dark matter in the outer parts of the 
galactic halos, but its extent is uncertain because of the limited 
range probed by observations.
More recent observational studies of rotational velocity 
in dwarf and low surface brightness galaxies have suggested
that the amount of dark 
matter in the central regions is smaller than predicted for the average 
galactic population by
CDM models (e.g. \citeNP{1994Natur.370..629M}, 
\citeNP{1994ApJ...427L...1F}, \citeNP{2000ApJ...543..704D}, 
\citeNP{2001ApJ...552L..23D}).  The CDM models
predict very concentrated dark matter haloes. This is usually parametrized
using a universal dark matter profile such as NFW \cite{1997ApJ...490..493N} 
\footnote{Although 
other profiles have been proposed that differ significantly from NFW
in the inner parts of the halo, they agree well with NFW in the 
outer parts \shortcite{2001ApJ...554..903K}. Since for the present work we are concerned predominantly
with the outer parts we only use NFW profile.}. 
For a given virial mass
of the halo, in this paper defined as the mass enclosed within a sphere of
radius $r_{200}$ within which
the density is 200 times critical density, the
profile has a characteristic shape which depends only on a single parameter.
For NFW profile, in which the slope continuously
changes from the inner value of -1 to the outer value of -3,
the free parameter is often
defined as the concentration $c_{200}$, which is the 
ratio between the radius $r_{200}$ and the scale radius
$r_s$ where the slope is close to $-2$. Higher concentration parameters
imply higher densities at the scale radius $r_s$. On galactic scales CDM
models predict $c_{200}\sim 7-15$ depending on the halo mass, 
matter density, shape and 
amplitude of the power spectrum. 
%In general, models with higher amplitude 
%of fluctuations, less negative effective slope or higher matter density 
%predict larger $c_200$ \cite{2001ApJ...554..114E}. 

However, observational case for low concentrations
is not clear and different
conclusions have been reached by other studies (e.g. 
\citeNP{2001MNRAS.325.1017V}). 
Even if the observational results of low dark matter density in the inner
regions are confirmed, they do not necessarily indicate a problem 
for the CDM models.
The variation
in the profile shapes between halos 
is large and it is not clear whether the dwarf or
low surface brightness galaxies, which show strongest evidence 
for low density cores, 
can be associated with the average 
halo population. 
For example, if these galaxies are associated with a 
population that had a major 
merger in the recent past, 
this may lead to a significantly flatter and less concentrated 
halo structure than the average population (e.g. \shortciteNP{2001astro.ph..8151W}). 
Another possibility is that astrophysical processes such as bar rotation 
redistribute the dark matter in the inner parts of the halo \cite{2001astro.ph.10632W}.
However, if the case of low concentrations is extended to the population 
of galaxies as a whole and to the outer parts of the halo, which cannot 
be affected by astrophysical processes, then the CDM crisis would become 
much more severe.

It would thus be useful to have some information on the halo structure
for the mean population of galaxies. While many observationally 
determined rotation velocity 
profiles exist, the main uncertainty has always been the relative 
contribution between the baryonic component in the gas, bulge/ellipsoid
and disk
and the dark matter in the halo. The baryonic and dark matter 
components are difficult 
to separate because the conversion from the star light (and, to a lesser
extent, gas density) to baryonic mass is still uncertain from the 
stellar population synthesis models and other studies.
The situation is further 
complicated by the fact that baryonic component is not dynamically 
negligible and during its condensation
it induces a response of the dark matter halo changing 
its original distribution in the inner parts. While these problems
are ameliorated for the low surface brightness galaxies, they are 
almost never negligible. 

Since the modelling of relative disk or spheroid and 
halo contributions is rather
difficult in the inner parts it is useful to concentrate on the 
outer parts of the halo, where the baryon influence is less important. 
However, optical 
rotation curves typically extend only out to  
$10-20$kpc and even the 
ones based on HI  measurements do not extend beyond 30-50kpc. 
Similarly, velocity dispersion studies in early type galaxies also 
do not extend past a few effective radii and even X-ray studies of
large ellipticals do not extend beyond 10 effective radii \cite{1999ApJ...518...50L}.
Over this limited range the data on disk $L_*$ galaxies 
indicate that the rotation curves rise
in the inner parts of the disk and then stay approximately 
constant or decline somewhat 
out to the outer limit of observations. This flatness indicates
that approximating the density profile as isothermal, $\rho(r) \propto r^{-2}$,
is a good approximation to the matter profile over this range. 
On the other hand, 
theoretical CDM models predict that the velocity profile of the
dark matter increases out 
to twice the scale radius $r_s$ 
and then slowly declines beyond that as the slope gradually decreases
towards -3. The fact that we do not observe a decrease in rotation 
velocity at radii below $2.15r_s \sim 30-50$kpc
is presumably due to the baryonic effects, which increase the mass in 
the center.
The decrease at large radii is however a robust prediction of these models. 
 
While, until recently, no accurate determination of the mass profile 
at large radii has been available, recent galaxy-galaxy lensing observations
by the SDSS team \shortcite{2001astro.ph..8013M} have improved the 
situation significantly by obtaining the morphology and luminosity 
dependence of the signal. Theoretical analysis of lensing data
must take into account not just galactic halos, but also those from groups 
and clusters, which dominate on large scales above 200-300$h^{-1}$kpc
\cite{gs02}.
These results show that a late type $L_*$ 
galaxy, with $L_I=2.7\times 10^{10}h^{-2}L_{\sun}$ (where we have applied
a 30\% internal extinction correction in $I$ band),
has a mass $M_{200}\sim (3.4\pm 2.1)\times 10^{11}h^{-1}M_{\sun}$, corresponding to the 
circular velocity at the virial radius $v_{200} \sim  115$km/s.
While the error is still quite large it is gaussian distributed in mass, so  
an increase in $v_{200}$ by 35\% to 150km/s
is excluded at 95\% confidence level.
This number can be 
compared to the maximum rotation velocity for such a galaxy, which is 
208km/s with a small scatter \shortcite{1997ApJ...477L...1G}, the well known Tully-Fisher relation. 
Comparing the two values shows that a decrease in rotation velocity from 
the optical to the virial radius indeed occurs for the average 
population of spiral galaxies. The decrease is large, almost a factor 
of 2 and even if we push the virial velocity up by 2-$\sigma$
to $v_{200}= 150$km/s the 
decrease is still around 40\%.
As shown in this paper, a similarly large decrease in rotation velocity 
is obtained also 
for early type galaxies which are not rotationally supported. 

These results are interesting, since they are in the direction predicted by 
the CDM models and demonstrate that the halo profiles indeed become 
steeper than -2 in the outer parts of the halo. 
They set tight limits
both on the structure formation models, 
by limiting the acceptable range of concentration parameters, and on the 
disk/spheroid formation models, by constraining the stellar mass to light 
ratio. Only for specific values of stellar mass to light ratio and 
halo concentration can one satisfy these constraints.
The purpose of this paper is to investigate the constraints in detail
using halo and disk/spheroid formation models. 

Previous work on this subject has 
explored the constraints from 
the rotation velocity at optical radius given by the zero point of
Tully-Fisher relation
(\citeNP{2000MNRAS.318..163M}, 
\citeNP{2000ApJ...538..477N}, \citeNP{2001ApJ...554..114E}). 
In the absence of virial mass information 
hese models must rely on additional assumptions
to derive the constraints on cosmological models. 
The advantage of the
additional information from lensing is that it provides another 
dynamical constraint at large radii, which can
remove some of the modelling 
uncertainties present in previous modelling. In addition, 
while previous work only explored the constraints from late type 
galaxies, in this paper we also investigate the constraints from 
the early type galaxies. We find these are more robust both because the 
virial masses are more accurately determined and because the velocity 
dispersions at optical radii are obtained from the same SDSS 
sample. 

\section{Late type galaxies}

The average rotation velocity at optical radii can be obtained from the 
Tully-Fisher (TF) relation, which in $I$-band is given by \shortcite{1997ApJ...477L...1G}
\begin{equation}
L_I=2.7\times 10^{10}\left({v_{\rm opt} \over 208{\rm km/s}}\right)^{3.1}h^{-2}L_{\sun},
\label{vrot}
\end{equation}
where we used $I-5\log h=-21-7.68[\log(2v_{\rm opt})-2.5]$ 
and $I_{\sun}=4.15$.
We denote with $v_{\rm opt}$ the maximum rotation 
velocity typically achieved at the optical radius (roughly 3 times the 
scale radius of the disk $R_d$, which is not to be confused by the 
scale radius of the halo $r_s$). At this radius the rotation curve still has 
a significant contribution from the disk. 
Rotation curves at larger radii show that, for this range of 
luminosities, the rotation curve at $L_*$ are flat or decline slightly
out to the largest radius observable, typically a few optical radii
(\citeNP{1991AJ....101.1231C},
\citeNP{1996MNRAS.281...27P}, 
\citeNP{2001astro.ph..8225V}).

We would like to compare the dark matter velocity in the inner 
parts of the halo to the 
SDSS galaxy-galaxy lensing results. For a late type 
$L_*$ galaxy with $i^*-5\log h =-21.26$
one finds $M_{200}=3.4\times 10^{11}h^{-2}M_{\sun}(c_{200}/10)^{-0.15}$ \cite{gs02}. 
Using $i^*_{\sun}=4.52$ \cite{2001AJ....121.2358B},
applying an average 0.3 magnitude internal extinction correction
\cite{2001astro.ph..8225V} 
and converting from the virial mass to the virial velocity using 
the relation $GM_{200}/r_{200}=v_{200}^2$ we find
\begin{equation}
L_I=2.7\times 10^{10}\left({v_{200} \over 115{\rm km/s}}\right)^3
\left({c_{200}\over 10}\right)^{-0.05}h^{-2}L_{\sun}.
\label{v200}
\end{equation}
The dynamical range of galaxy-galaxy lensing is still rather small 
and for late type galaxies luminosity dependence of the virial mass
cannot be established using the present sample, so we focus on $L_*$ galaxies,
which dominate the late galaxy galaxy-galaxy (g-g) lensing signal. 

By combining the two equations above
we can determine the best fitted ratio 
\begin{equation}
{v_{\rm opt} \over v_{200}} \sim 1.8.  
\label{vrat}
\end{equation} 
Since the virial mass of late type galaxies is consistent with 0 at 
2-$\sigma$ level there is no upper limit to the velocity ratio, while 
the lower limit is given by $v_{\rm opt}/v_{200}>1.4$ at 2-$\sigma$,  
where the error budget 
includes the statistical errors on the zero point of Tully-Fisher relation 
\shortcite{1997ApJ...477L...1G}, but is dominated by the error 
on the virial mass from galaxy-galaxy lensing \cite{gs02} 
and we ignored the small concentration dependence. 
In addition to the statistical errors there are also possible 
systematic differences between the different Tully-Fisher
zero point determinations (\shortciteNP{1997ApJ...477L...1G}, 
\shortciteNP{1996ApJ...457..460W}, \citeNP{1997MNRAS.290L..77S}) which can be up to 0.2 magnitude. 
Similarly there could be systematic differences between color selected 
late type galaxies
\cite{2001astro.ph..8013M} and those selected for
rotation velocity studies, although morphological studies indicate that
the late type sample is 
dominated by Sb/Sc morphological type used also in TF studies 
\cite{2001astro.ph..7201S}. 

What constraints does equation \ref{vrat} imply on the structure formation 
models? As discussed above, a decrease in rotation velocity implies that the 
mass profile is steeper than isothermal in the outer parts of the 
halo. This can be either because the dark matter profiles are steep or 
because the stellar disk has a significant 
contribution to the rotation velocity (or both).  
An NFW profile is given by 
$\rho(x)=\rho_s  x^{-1}(1+x)^{-2}$, where $x=r/r_{s}$.
The rotation velocity is $v_c(r)^2=GM(r)/r$, where 
$M(r)=4\pi \rho_sr_s^3A(c)$ and $A(c)=\ln(1+c)-c/(1+c)$. For NFW profile the rotation velocity 
increases at small $x$ up to 
the peak at $x=2.16$ and then declines gradually to the virial radius. 
The ratio between the maximum and virial velocity is 
$v_{\rm max} / v_{200}=0.46[c/A(c)]^{1/2}$
\cite{2001MNRAS.321..559B}.
For the range of concentration parameters predicted by CDM models 
($c<20$) one finds $v_{\rm opt} / v_{200}<1.4$.
It is thus unlikely that the dark matter halo can explain the decrease
by itself, since 
even for very concentrated halos the velocity decrease is less than observed. 
We must therefore include the contribution from the disk.

We will use a model with a thin exponential 
disk in a dark matter halo (\citeNP{1998MNRAS.295..319M}, 
\citeNP{2000ApJ...530..177V}). 
We model the disk as a thin exponential surface density 
profile,
\begin{equation}
\Sigma(R)=\Sigma_0\exp(-R/R_d),
\end{equation}
with the total disk mass given by $M_d=2\pi \Sigma_0R_d^2$. 
The disk mass is related to the disk luminosity using disk 
mass to light ratio $\Upsilon_I=M/L$. We will neglect 
the bulge contribution both to the luminosity 
and to the rotation curve (see \shortciteNP{1998MNRAS.295..319M}  
for a discussion of this assumption).  
We will use observations to provide 
the typical scale length $R_{d}$ 
of the galactic disks. For an $L_*$ galaxy observations give $R_d\sim 
3.5h^{-1}$kpc (\citeNP{1997AJ....114.2402C}, \citeNP{1996A&A...313...45D}). 
The scatter for this quantity at a given luminosity 
is rather large, since galaxies come with a range of surface brightnesses.
Theoretically, the scatter has been 
linked to the scatter in the spin parameter of haloes 
\shortcite{1998MNRAS.295..319M}. 
However, since in this paper we are primarily concerned with 
the average properties of galaxies and not in the scatter around 
the mean
we will assume $R_d= 3.5h^{-1}$kpc in the analysis as a mean value
for a typical $L_*$ galaxy. 

Disk gravity contributes to the measured rotation velocity. 
In addition, disk gravity also induces a response of the dark 
matter halo in the inner regions. The standard approach to 
model this is to assume an adiabatic contraction of the halo, 
which remains spherical, so that the angular momentum of individual 
particles is conserved (see \citeNP{1984MNRAS.211..753B}, \shortciteNP{1986ApJ...301...27B} and \citeNP{2000ApJ...538..477N} 
for more details and numerical tests on
the validity of this model). This leads to an implicit equation for the
final radius $r_{\rm f}$ of dark matter mass as a function of initial 
radius $r_{\rm i}$
\begin{equation}
M_{\rm DM}(r_{\rm i})r_{\rm i}=[M_{\rm DM}(r_{\rm i})(1-f_*)+M_{\rm s}(r_{\rm f})]r_{\rm f},
\end{equation}
where $M_{\rm DM}(r)$ and $M_{\rm s}(r)$ are dark matter and stellar mass, 
respectively and $f_*$ is the stellar mass fraction of the halo.
Assuming that the baryons which do not end up in the disk have the same
distribution as the dark matter allows one to solve the system completely for 
a given disk mass and scale length and for a given halo profile.  
Flattened nature of the disk is used when obtaining stellar rotation 
velocity from the mass profile \cite{BT87}.
As a free parameter we will use the stellar mass to light ratio $\Upsilon_I$ (expressed 
in solar units), which 
from the known virial mass and luminosity of $L_*$ galaxy (and ignoring 
bulge contribution to the luminosity) can be 
related to $f_*$ as
\begin{equation}
\Upsilon_I=M_d/L_*={3.4\times 10^{11}h^{-1}M_{\sun}f_* \over 2.7\times 10^{10}h^{-2}L_{\sun}}
\approx 12f_* {hM_{\sun} \over L_{\sun}}.
\end{equation}
Typical values are $\Upsilon_I=(1-2)h$ (e.g. \citeNP{1997A&A...328..517B}), giving
$f_*=0.1-0.2$. Note that $f_*$ should not exceed $\Omega_b/\Omega_m \sim 0.04/\Omega_m$, since only the 
baryons within the virial radius can condense to make stars. This implies $\Omega_m<0.4$ for the fiducial value of $M_*$ and $\Omega_m<1$ for the 95 \% c.l.
on $M_*$. This is in agreement with other determinations that give 
$\Omega_m<0.4$, so 
the fraction of baryons converted to stars does not 
exceed the available supply, although it comes quite close to this limit and 
is an argument against high values for $\Upsilon_I$.

\begin{figure}
\begin{center}
\leavevmode
\epsfxsize=3.0in \epsfbox{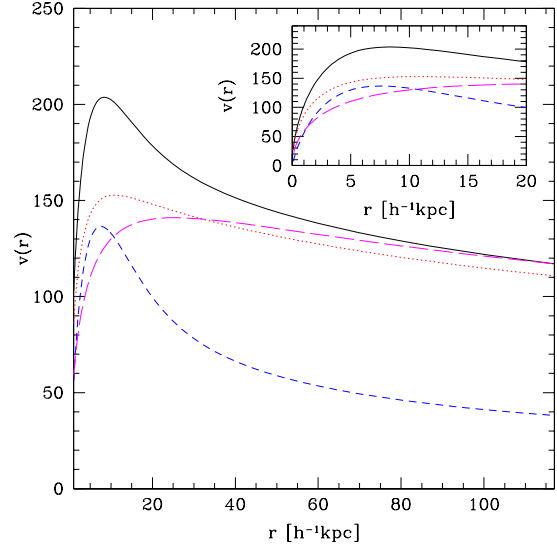}
\end{center}
\caption{Rotation velocity contributions from the disk (short dashed), 
dark matter (dotted) and total (solid) out to the virial radius. 
Also shown is the 
dark matter without adiabatic response to baryon contraction (long dashed).
Insert plots the same over the radii accessible with HI data, 
showing that the rotation curve is close to flat over this range.
We do not model the bulge/bar, so the predictions within inner few kpc
are not reliable.
}
\label{fig1}
\end{figure}

We can solve for $v_{\rm opt}/v_{200}$ for any given $f_*$ (or $\Upsilon_I$) 
and $c_{200}$. 
An example with $\Upsilon_I=1.7h$ and $c_{200}=12$ is shown in figure \ref{fig1}. 
One can see that the velocity ratio of 1.8 can be naturally
obtained in a model with 
concentration parameter in the range predicted by CDM models, $8<c_{200}<15$ 
 (\citeNP{2001ApJ...554..114E}, \shortciteNP{2001MNRAS.321..559B}), and 
with the expected stellar mass to light ratio, $1h<\Upsilon_I<2h$. 
The rotation velocity is reasonably 
close to flat over the optical region,
but decreases 
by 10-20\% out to the largest range observable in HI, 
in agreement with the observations for this range of 
luminosities (\citeNP{1991AJ....101.1231C}, 
\citeNP{2001astro.ph..8225V}). 
The disk and dark matter contributions to the mass within the optical radius 
are comparable, each contributing 50\% in this example. This implies
that this model 
satisfies the requirement that
the zero point of TF relation is independent of the disk surface brightness,
which requires about
50\% dark matter contribution to the rotation velocity at optical radius
\cite{2000MNRAS.318..163M}.
Note that the adiabatic response 
of dark matter is quite significant and dark matter 
would be subdominant if it were not 
compressed by baryonic condensation. 
Because of this the
dark matter contribution to the rotation curve is never negligible, 
even in the inner parts of the galaxy. However, our predictions may
not be reliable inside the optical radius, where bulge makes a 
significant contribution and adibatic approximation may not be 
valid \cite{2001astro.ph.10632W}.

\begin{figure}
\begin{center}
\leavevmode
\epsfxsize=3.0in \epsfbox{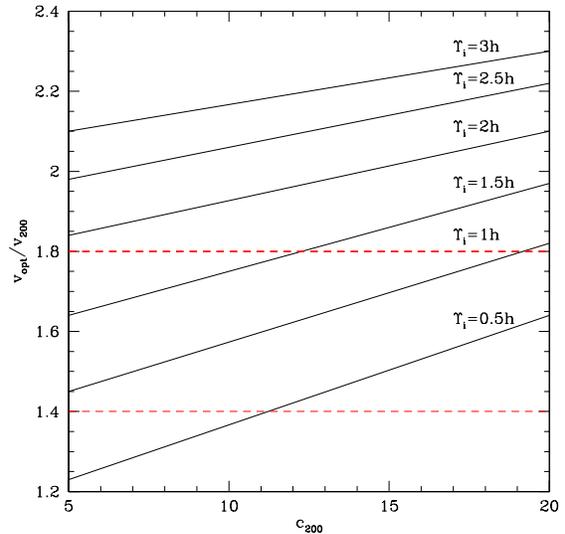}
\end{center}
\caption{
Lines of constant $\Upsilon_I$ (times $h$) 
as a function of $c$ versus
$v_{\rm opt}/v_{200}$. Thick dashed horizontal line 
is the observationally
determined value $v_{\rm opt}/v_{200}=1.8 $.
A 2-$\sigma$ (98\% confidence level) 
lower limit is $v_{\rm opt}/v_{200}>1.4$ (thin dashed line), while currently
the data provide
no upper limit since the virial masses are consistent with 0.
}
\label{fig2}
\end{figure}

The example above was chosen based on the typical values for concentration and stellar 
mass to light ratio. More generally for any choice of one there will be a particular value 
for the other that satisfies the constraint in equation 
(\ref{vrat}). This is shown in figure \ref{fig2} for 
a family of stellar mass to light ratios. 
Low values of $\Upsilon_I$ cannot be made compatible with the constraint 
in equation \ref{vrat}, 
unless the concentrations are unreasonably high (e.g. $c>20$ for 
$\Upsilon_I<1h$). 
Low concentrations are also not acceptable unless
the observed $v_{\rm opt}/v_{200}$ is decreased by 
2-$\sigma$ to 1.4. In this case  
one must adopt either 
very low concentrations for reasonable $\Upsilon_I \sim 1.5h$ or very low 
$\Upsilon_I<0.7h$ for reasonable $c \sim 10$. On the other hand, a positive 
deviation in $v_{\rm opt}/v_{200}$ can be explained by using a somewhat higher 
values for $\Upsilon_I$.
We conclude that the results are just what is expected 
from the CDM models with standard concentrations and standard stellar 
mass to light ratios. Only if the virial masses deviate by 2-$\sigma$
in the positive direction
from the mean value does one run into problems with the standard stellar mass 
to light ratios and one requires $\Upsilon_I=0.5h$. 
The majority of late type galaxies  
therefore show no 
evidence for shallow density profiles in the outer parts of the halo. 

\section{Early type galaxies}

Early type galaxies have some advantages  
in constraining the outer parts of the halo profiles. 
The main advantage is that they
show a stronger g-g lensing signal, so that both $M_*$ and 
its scaling with luminosity are reliably determined from the current data 
and the errors associated with it are significantly smaller.
Moreover, one can study their early 
type dynamics using the same spectroscopic SDSS sample also used for g-g lensing
(the actual samples used in the analysis here 
are not completely equal since the analysis for the
early type dynamics in \citeN{2001astro.ph.10344B} was using a larger sample than the lensing 
analysis in \citeN{2001astro.ph..8013M}, but the
statistical properties of the two samples should be very similar). The 
fundamental plane relations as derived by 
\shortciteN{2001astro.ph.10344B} are $L \propto \sigma^4 \propto R_{\rm e}^{1.5}$, where $\sigma$ is the 
central stellar velocity dispersion and $R_{\rm e}$ is the effective 
radius of de Vauculeurs profile. The values at $L_*$ in $i'=-21.26$ are 
$\sigma=177$km/s and $R_{\rm e}=2.7h^{-1}$kpc (note that we are using $L_*$ for 
the luminosity function of the whole sample given in 
\shortciteNP{2001AJ....121.2358B}, 
not just the early type).

For early type galaxies 
the conversion from the stellar velocity dispersion to matter circular velocity 
is less straightforward. In principle one can obtain it by solving
the Jeans equation, which however depends on the unknown anisotropy of velocity
dispersion. 
If the circular velocity of the matter does not 
change very much over the optical region then virial theorem guarantees that 
$v_{\rm opt}=3^{1/2}\langle \sigma \rangle$, if $\langle \sigma \rangle$ 
is luminosity weighted line of sight 
velocity dispersion.
In practice luminosity average is difficult to achieve and the velocity 
dispersion decreases with radius, so more often observers report the
central velocity dispersion $\sigma_{\rm central}$. Based on detailed 
kinematic analysis and on comparison between strong lensing and 
stellar dispersions \citeN{1994ApJ...436...56K} argues
that in this case the relation is closer to $v_{\rm opt}=2^{1/2}\sigma_{\rm central}$.
In SDSS the aperture is determined by SDSS fibers and is $r_{\rm fiber}=1.5"$, 
implying 
that for nearby galaxies only central parts of the galaxy are detected, 
while for distant galaxies most of the light is observed. \citeN{2001astro.ph.10344B} 
attempt to correct for this using an empirical fit, $\sigma_{\rm central}/
\sigma=(8r_{\rm fiber}/R_{\rm e})^{0.04}$,
to obtain the central velocity dispersion. The correction is empirical and
appears
to be somewhat small to account for the suggested 22\% difference between the 
central and luminosity weighted velocity dispersion, since one needs
very small $R_{\rm e}=0.15"$ to achieve this. We will use an intermediate
value of $v_{\rm opt}=1.5\sigma$ to convert from the central velocity 
dispersion to the rotation velocity at the optical radius. This
should have at most a 10\% systematic uncertainty attached to it.
This conversion
agrees well with the studies of slowly rotating elliptical 
galaxies where both $v_{\rm opt}$ and $\sigma_{\rm central}$ have been
measured \cite{2001AJ....121.1936G}. 

Another ingredient in the modelling is the dark matter response to 
baryonic contraction. 
We model it again using the adiabatic model. 
In principle there is no reason why such a model would be appropriate 
for elliptical galaxies, where 
stars are not on circular orbits, but numerical 
simulations of galaxy formation have found that adiabatic compression model
works remarkably well even for such systems \cite{gotbrath02}, perhaps
as a consequence of conservation of radial action in such systems. 
We use the Hernquist 
profile \cite{1990ApJ...356..359H}, which
has been shown to give a light profile very close 
to the de Vaucouleurs profile and has an analytic 3-d radial distribution,
to model the star distribution \cite{2001ApJ...561...46K}.
Our canonic value for the stellar mass to light ratio in $i'$ is 
a factor of 2
higher than that of late type galaxies in the same band,
$\Upsilon_{i'}=3h M_{\sun}/L_{\sun}$. 
This is based on $K-i'$ color difference between early and late type 
galaxies in SDSS, which is around 0.2-0.3 magnitudes (without
internal extinction correction, \shortciteNP{2001astro.ph.11024I}) and 
the fact 
that in K band luminosity to stellar mass conversion only depends 
on the age of population and differs by less than a factor of 2 between 
early and late type galaxies for reasonable IMF and assuming ages above
3Gyr 
\shortcite{2002astro.ph..1207D}.  
This stellar mass to light ratio is again rather uncertain and significantly 
higher values have been suggested in the literature. Most of the direct 
studies are done in B band, but even after correcting for a factor of 2-3 
difference between B and i' luminosity for early type galaxies it is on 
the low side based on the dynamical studies of central regions using 
the minimal halo models \shortcite{2001AJ....121.1936G}. The simplest
explanation is that part of the mass is actually due to dark matter, as 
discussed further below.
Note that for $\Upsilon_{i'}=4h M_{\sun}/L_{\sun}$ the stellar to 
virial halo mass ratio for early and late type galaxies become equal and are
approaching the maximal baryon to dark matter ratio still allowed by 
the observations. It is thus unlikely that the average stellar mass to light 
ratio can significantly exceed this value if the virial masses
from g-g lensing are correct.

At $L_*=2\times 10^{10}h^{-2}L_{\sun}$  
the virial mass for early type galaxy is 
$(9.3 \pm 2.2)\times 10^{11}h^{-1}M_{\sun}$ \cite{gs02}, which translates to 
$v_{200}=(160 \pm 15)$ km/s
at $r_{200}=160h^{-1}$kpc. At the effective radius 
$R_{\rm e}=2.7h^{-1}$kpc the rotation velocity from optical velocity 
dispersion is
$v_{\rm opt} \sim 1.5\times 177{\rm km/s}=
265$km/s with a small error \cite{2001astro.ph.10344B}, leading to
\begin{equation}
{v_{opt} \over v_{200}} =1.68 \pm 0.2.
\end{equation}
The error is dominated by stellar to 
dark matter velocity dispersion conversion and  
virial mass uncertainty.
Here again we have a very large
decrease from the optical to the virial radius, which is inconsistent
with the flat rotation curve at more than 3-$\sigma$ level. This decrease is 
similarly large to the one observed for the late type galaxies. 
Such a 
decrease cannot be explained by the dark matter alone unless halos are 
extremely concentrated. More realistic models must include baryons, 
which make 
a significant contribution to the rotation velocity at optical radii, 
both by direct contribution and by compressing the dark matter. 

\begin{figure}
\begin{center}
\leavevmode
\epsfxsize=3.0in \epsfbox{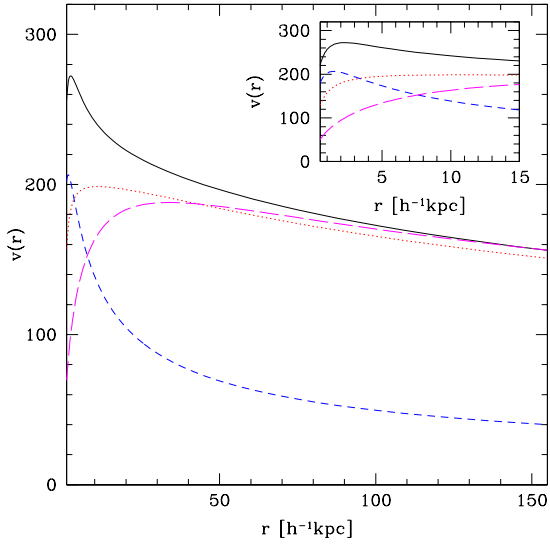}
\end{center}
\caption{Rotation velocity contributions from the spheroid (short dashed), 
dark matter (dotted) and total (solid) for an $L_*$ early type 
galaxy. Also shown is the 
dark matter without adiabatic response to baryon contraction (long dashed).
Insert plots the same over the radii accessible with optical data, 
showing that the profile is close to isothermal over this range.
}
\label{fig3}
\end{figure}

The resulting velocity profiles are shown in top of figure \ref{fig3} 
for the canonic values $\Upsilon_I=3h$ and $c_{200}=10$ (we use 
a somewhat lower $c_{200}$ than for late types 
since concentration is expected to decrease with halo mass). 
The maximum rotation velocity peaks very close to the optical radius and 
has a value of $v_{max}=270$km/s. This is in a close agreement 
with the observed value $v_{\rm opt}=265$km/s and is well within the 
estimated error of 30km/s, indicating that this model has 
no problem explaining the observed ratio of
the optical to virial rotation velocity. Note that 
at the optical radius the baryon and dark matter contributions are 
comparable, while at somewhat 
larger radii dark matter dominates. The resulting profile is much closer
to a constant velocity 
SIS profile than if just light was contributing to the mass. 
This is in a good agreement with the conclusions from strong 
lensing \cite{1995ApJ...445..559K}, rotation velocity studies of ellipticals 
\cite{2001AJ....121.1936G} and X-ray studies of ellipticals \cite{1999ApJ...518...50L}, which indicate that 
flat rotation curve over the optical region ($r<10h^{-1}$kpc)
is a better fit to the data than a constant stellar mass to light 
ratio with no dark matter contribution. 

\begin{figure}
\begin{center}
\leavevmode
\epsfxsize=3.0in \epsfbox{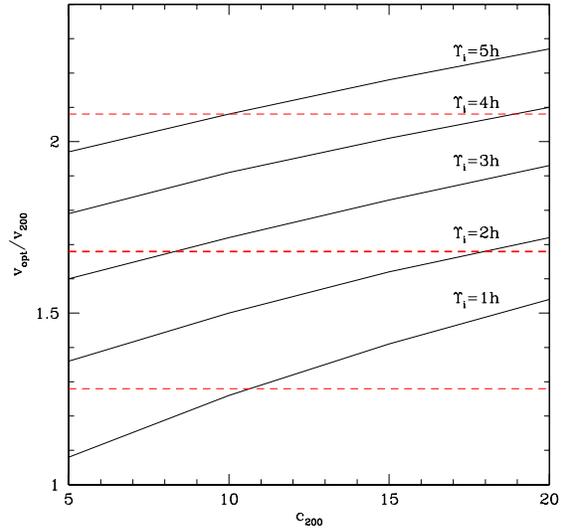}
\end{center}
\caption{
Same as figure \ref{fig2} for early type galaxies at $L_*$. Also shown 
(thick dashed) is the best 
fitted value $v_{\rm opt}/v_{200}=1.68$ together with 
2-$\sigma$ (95\% confidence level) lower and upper limits (thin dashed). 
}
\label{fig6}
\end{figure}

A general exploration of $c$ versus $\Upsilon_i$ is 
shown in figure \ref{fig6} for $L_*$. Within 1-$\sigma$ of the best fitted value 
for $v_{\rm opt}/v_{200}$ one has $2h<\Upsilon_i<4h$ assuming $c=10$, 
whereas at 2-$\sigma$ level this is extended to $1h<\Upsilon_i<5h$.
While the range of stellar mass to light ratios suggested in the 
literature is rather large and extends even above $\Upsilon_i>5h$, such 
high values are typically found for minimum halo models and are 
thus an upper limit. Our analysis suggests that $\Upsilon_i<5h$
both because of the dynamical constraint and because of the limited baryon 
supply. Very high values of stellar mass to light ratios are also 
not compatible with the stellar population synthesis models using the 
observed IMF (e.g. 
\citeNP{1998MNRAS.294..705K}).  

Since the scaling of virial velocity with luminosity $L\propto v_{200}^{2.5}$ 
\cite{gs02}
differs from the scaling
of optical velocity with luminosity $L \propto v_{\rm opt}^4$
above $L_*$ we must also compare at 
a higher luminosity. We choose $L=7L_*$, which corresponds to the 
highest luminosity bin in SDSS g-g lensing analysis \shortcite{2001astro.ph..8013M} and is dominated by
early type galaxies. At this luminosity
one finds $M_{200} \sim 10^{13}h^{-1}M_{\sun}$ \cite{gs02}, which 
corresponds to $v_{200}=(310 \pm 30)$km/s. 
The corresponding central velocity dispersion is
$\sigma_{\rm central}=290$km/s, implying $v_{\rm opt}=435$km/s
and $R_{\rm e}=10h^{-1}$kpc \cite{2001astro.ph.10344B}. Here the ratio is 
$v_{max} / v_{200} =1.4 \pm 0.2$.
The dominating error is the virial mass at this luminosity. 

\begin{figure}
\begin{center}
\leavevmode
\epsfxsize=3.0in \epsfbox{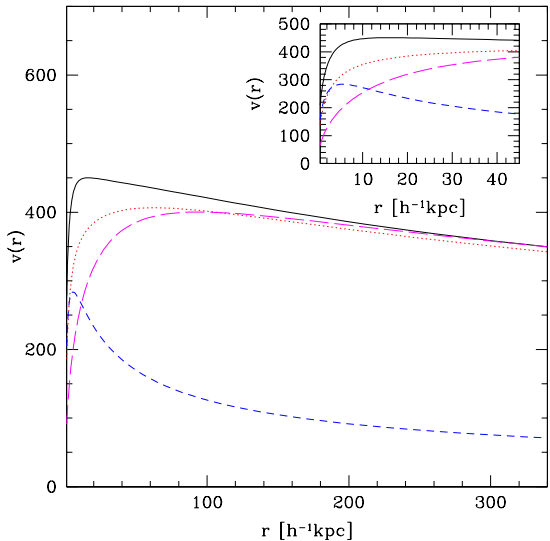}
\end{center}
\caption{Same as figure \ref{fig3} for 7$L_*$ early type galaxy.
}
\label{fig4}
\end{figure}

Results of the adiabatic model calculation are shown in figure \ref{fig4} 
using $\Upsilon_{i'}=3h$. 
We have used an even lower concentration value $c=8$ for this 
higher mass halo, since 
numerical simulations find $c \propto M^{-0.14}$ \cite{2001MNRAS.321..559B}. 
We again find the rotation velocity at the optical radius 
exceeding that at the virial radius, but the excess is significantly 
smaller now. 
The rotation curve is very flat for $r>5h^{-1}$kpc with the value 
at the optical radius
around 445km/s, in good agreement with the observed value of 435km/s. 
The peak value of the dark matter rotation curve alone is around 400km/s
and is close to the value at the optical radius 
given that the rotation curve is so flat.
Note that at this luminosity the dark matter dominates already at the 
optical radius and it is only below 5$h^{-1}$kpc that baryon mass
exceeds that of the dark matter. However, the baryons do have a significant 
effect on the dark matter through the adiabatic compression, so that 
even though the dark matter dominates at the optical radius it would 
have been comparable to baryons if there was no dark matter compression. 
Our analysis suggests 
that the data are consistent with 
brighter ellipticals being more dark matter dominated at the
optical radii than the fainter ones and the stellar mass to light 
ratio not varying with luminosity.

\begin{figure}
\begin{center}
\leavevmode
\epsfxsize=3.0in \epsfbox{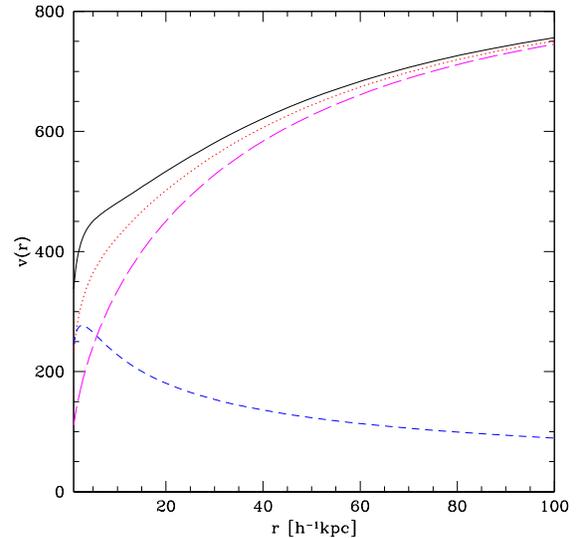}
\end{center}
\caption{Same as figure \ref{fig3} for $2\times 10^{11}h^{-1}M_{\sun}$
early type galaxy in 
a $2\times 10^{14}h^{-1}M_{\sun}$ halo as a model for M87/Virgo.
}
\label{fig5}
\end{figure}

It is worth noting that the model can be extended by yet 
another order of magnitude in virial mass, to cluster masses
with an elliptical cD galaxy at the 
center. A typical example is Virgo cluster with M87 at the center. 
For M87/Virgo the rotation velocity 
increases from the optical radius $r_{\rm E}\sim 5h^{-1}$kpc 
value $v_{\rm opt}\sim 450$km/s \cite{2001ApJ...553..722R}, 
to around 950km/s deduced from the observed 
X-ray temperature around 3keV. 
An example that can reproduce these constraints is shown in 
figure \ref{fig5}, where a halo with $M_{200}=2\times 10^{14}h^{-1}M_{\sun}$,
$c=6$ and spheroid with
$M_{\rm stellar}=2\times 10^{11}h^{-1}M_{\sun}$ have been used 
\cite{1999ApJ...524L..15G}. In this case
one has a ratio in 
rotation velocity from optical to virial radius $v_{\rm opt}/v_{200}\sim 0.5$. 
This is caused primarily by the cluster halo profile 
becoming shallower than $r^{-2}$ in the inner parts. While baryons 
in cD
increase the rotation velocity from the pure dark matter value in the inner 
parts of the halo they cannot 
pull in outer parts of the dark matter halo, which remain unaffected. 
The trend of optical to virial rotation velocity 
ratio decreasing with halo mass observed in elliptical galaxies
is therefore expected as a consequence of more 
massive halos being less concentrated and less steep than isothermal in 
the central region, although for quantitative predictions baryons and 
their effect on dark matter must also be included. 

\section{Discussion}

In this paper we compare average properties of
rotation velocities between optical and virial radii based on 
optical and galaxy-galay lensing measurements, respectively. 
Our model is statistical in nature, since 
it is not based on analysis of individual galaxies, but on their
average properties as a function of luminosity. On the other hand, 
the galaxies in our sample were not chosen on the basis of any 
selection criteria, so our results should apply 
to the galaxy population as a whole. Moreover, the large dynamical range 
between optical region (of order a few kpc) and virial region 
(of order a few hundred kpc) allows one to measure slow changes in rotation 
velocity which are not possible to detect from each of the observations
individually.

Our main conclusion is that the rotation velocity 
in galaxies 
decreases significantly from optical radii to virial radii. 
Such a decrease is theoretically expected since
the dark matter profiles in CDM models are steeper than 
isothermal at large radii and contribution from 
stars further increases the rotation 
velocity at the optical radius. It has however 
not been clearly demonstrated previously 
because of a narrow range of scales observed. This demostrates the power of 
g-g lensing measuring the mass at large radii 
combined with the more traditional methods that measure mass at 
smaller radii (see also \citeNP{2001ApJ...555..572W} for 
a similar conclusion for early type galaxies at higher redshifts).
The concentration parameters obtained using reasonable stellar mass to 
light ratios are around $c\sim 10$ and are
in a good agreement with predictions of
$\Lambda$CDM model with $\Omega_m=0.3$ and $\sigma_8=0.8$, which
gives $c_{200} \sim 10$ at $M \sim 10^{12}h^{-1}
M_{\sun}$ \shortcite{2001ApJ...554..114E}. 
We find no evidence for low halo concentrations in the main galaxy 
population both for early and late type galaxies.
 
The theoretically predicted 
stellar mass to light ratios have considerable theoretical uncertainties.
Assuming CDM halo profiles and concentrations we find $\Upsilon_I \sim 
1.5h$ for late type galaxies and $\Upsilon_i \sim 3h$ for early type 
galaxies. These values are in a good agreement with the stellar population 
synthesis models (e.g. \citeNP{1998MNRAS.294..705K}, 
\shortciteNP{2002astro.ph..1207D}). While higher stellar mass to light ratios are often 
quoted in the literature (e.g. \shortciteNP{2001AJ....121.1936G}) these are based on minimal halo 
models and are thus often an upper limit. It is interesting to note that 
in our models the dark matter contribution to rotation velocity is comparable
to that of stars even below effective radius. This is because of the
adiabatic response, which compresses the dark matter in the center. While 
this model may be oversimplified it nevertheless suggests  
that it may be difficult or
impossible to separate the two components on the 
basis of dynamical studies. Using the minimal halo assumption the
stellar mass to light ratio in early type galaxies may be overestimated
up to a factor of two. 

While the optical to virial velocity ratio
is above unity both for early and late type galaxies, 
it is decreasing with halo mass from 1.8 at $3\times 10^{11}h^{-1}M_{\sun}$
to 1.4 at $10^{13}h^{-1}M_{\sun}$ (this statement is valid only for galaxies
above $L_*$ and it is possible that the trend is reversed towards the
low luminosity galaxies). This trend continues further into the 
cluster halo masses, where the ratio falls below unity. 
Such a trend is expected in models where halo profiles are less 
concentrated for higher halo masses, implying that the turnaround 
from an increase to a decrease in rotation velocity occurs at a larger 
radius relative to the virial radius. In addition, in more massive halos 
stars play a less important role both as a direct contribution to 
the rotation velocity and through their effect on the
dark matter.
The rotation velocity-luminosity scalings at optical 
radii, such as Tully-Fisher and Faber-Jackson relations, 
are not directly related to 
the properties of dark matter, but also require a proper modelling of
baryons and dark matter response to baryonic contraction (see also a
related discussion in \shortciteNP{2000ApJ...528..145G}, \citeNP{2001astro.ph..8160K} and \citeNP{2001astro.ph.12566V}).
While there are still uncertainties in the modelling 
of these processes, the simple models presented here reproduce
well the constraints from the data both for early and late 
type galaxies. 

How do our results compare to previous work?
The zero point of TF relation problem (\shortciteNP{2001ApJ...554..114E}, 
\citeNP{2000MNRAS.318..163M})  
is the closest to the TF
analysis done here. In the absence of virial mass information
the value of rotation velocity at a given luminosity 
does not suffice to make any general conclusions, so
in general one has to make additional 
assumptions and/or modelling. 
For example, the stellar mass fraction in the halo $f_*$ obtained 
in previous work was lower, which lead to a higher virial mass 
for a given luminosity, 
which in turn requires lower concentrations and/or stellar mass to light 
ratios. By increasing $f_*$ close to its maximum value 
the virial mass can be reduced and this 
alleviates the problem. The same solution also solves the suggested 
overprediction of dark matter at the solar radius in our own galaxy
\cite{2001ApJ...554..114E}, 
since again if the stellar fraction is higher the virial mass
can be lower (there may be
additional problems for CDM profiles in the inner parts of our galaxy; 
e.g. \citeNP{2001MNRAS.327L..27B}).

For early type galaxies 
it has been suggested that the concentrations 
are low from the strong lens statistics \cite{2001ApJ...561...46K}, since very 
concentrated halos would overpredict the expected number of lenses. This is 
a difficult method to use since the expected number of lenses is very 
sensitive to the assumed luminosity function
for early type galaxies as a function of redshift, which still has
considerable uncertainty. Additional uncertainty arises from the 
adopted values for stellar mass to light ratios, which again can change 
the lensing statistics significantly. The lensing results are 
still compatible with low density CDM models, suggesting 
there is no discrepancy with our results, although more work is required to 
study this in detail. 

There are other problems that have been suggested as troublesome for CDM, 
such as detailed shapes of velocity profiles in the optical region 
\cite{2001ApJ...552L..23D}, bar rotation \cite{2000ApJ...543..704D}
 or the halo 
structure of the Milky Way \cite{2001MNRAS.327L..27B}. These probe inner 
regions of the galaxy where complicated physical processes may be 
taking place, so there is considerable more uncertainty in their 
theoretical predictions. For example, 
there are processes such as bar rotation that can disrupt dark matter 
cusps \cite{2001astro.ph.10632W}. 
It has recently been shown that CDM profiles can 
fit most of the rotation curves for normal galaxies \cite{2002astro.ph..1352J}. 
The galaxies that appear to be a problem for CDM belong to 
one of the specific subsamples, 
such as low surface brightness, dwarf or barred galaxies. 
It is possible, although not necessarily easy to arrange, that
these samples are qualitatively different from the main 
population, for example by forming later and thus being less concentrated. 
Yet another possibility is that problems arise only 
below $L_*$, since our analysis is only valid for galaxies around 
and above $L_*$. Clearly, more 
work is required to resolve these issues. However,
if the g-g lensing masses are correct, then for the main population of galaxies
around and above $L_*$
the CDM model predictions for the
amount of dark matter outside the inner few kpc 
do not exceed the observations, 
suggesting that the problems for CDM may not be as fundamental as previously 
suggested.

The author acknowledges the support of NASA, David and Lucille 
Packard Foundation and Alfred P. Sloan Foundation. 
I thank Mariangela Bernardi, \v Zeljko Ivezi\' c, Guinevere Kauffmann, Ravi 
Sheth and
Tommaso Treu for useful discussions and Matthias Steinmetz for providing 
their manuscript prior to publication.

     \bibliography{apjmnemonic,cosmo,cosmo_preprints}
	\bibliographystyle{mnras}

\end{document}